\documentclass[letter]{ptptex}

\usepackage{graphicx}
\usepackage{wrapft}

\def\picbox#1#2{\fbox{\vbox to#2{\hbox to#1{}}}}
\def\bra#1{\langle#1|}
\def\ket#1{|#1\rangle}

\def\scalar#1#2{\langle#1|#2\rangle}

\def\mean#1{\overline{#1}}
\def\ave#1{\left\langle #1\right\rangle}

\title{%
Value statistics of chaotic Wigner function
}

\author{
Martin \textsc{Horvat}\footnote{martin@fiz.uni-lj.si} and
Toma\v z \textsc{Prosen}\footnote{prosen@fiz.uni-lj.si}
}

\inst{%
Faculty of mathematics and physics, University of Ljubljana, Jadranska 19, SI-1000 Ljubljana, Slovenia
}

\abst{%
We study Wigner function value statistics of classically chaotic
quantum maps on compact 2D phase space. We show that the Wigner function
statistics of a random state is a Gaussian, with the mean value becoming
negligible compared to the width in the semi-classical limit.
Using numerical example of quantized sawtooth map we demonstrate that the
relaxation of time-dependent Wigner function statistics, starting from a 
coherent initial state, takes place on a logarithmically short 
($\propto \log\hbar$) time scale.}

\begin{document}

\maketitle
The Wigner function\cite{wigner} (WF) is an essential concept of phase space representation of quantum mechanics, namely it is a useful and faithful representation of a pure or mixed quantum state in terms of functions of canonical classical phase space variables. It has many applications in various branches of physics, in particular in quantum optics. However, WF cannot be interpreted as quantum phase space distribution as it can develop negative values, in particular due to well known oscillatory interference fringes following e.g. coherent wave-packet superpositions \cite{zurek}.
WF played an important role in the realm of quantum chaology. \cite{berry2,voros} \ In particular, it has been conjectured that WF of stationary 
eigenstates of bounded dynamical systems in quasi-classical regimes
condense, within classical scales, onto classically invariant components of 
phase space. On the other hand, 
within a smaller - quantum scale, e.g. of Planck cell size, the phase space structure of WF of typical, 
say random or ergodic states, is very much unknown. This question is important for the understanding of 
decoherence and quantum stability with respect to system's perturbations as discussed recently. 
\cite{zurek,PZ02,zurek2,Ben}\par
In this paper we address the question of the structure of typical WF from a statistical point of view. We define and analyze the statistical distribution of values of WF of a given quantum state. In particular, we are interested in the relaxation of this distribution with time, when we start from initial coherent state, and in the corresponding time scale. Here we limit ourselves on classically chaotic discrete time systems, namely chaotic quantum maps, where the full phase space is the only topologically transitive classical ergodic component.\par
We show that for fully chaotic quantum maps the limiting WF value distribution is a Gaussian. The average value is fixed by normalization, while the second moment (or the variance) diverges in the quasi-classical limit ($\hbar\to 0$), so we have roughly a symmetric distribution of positive and negative values of WF for random states. In addition, we show that for chaotic systems, the relaxation to equilibrium in a statistics of WF happens on a short $\propto\log\hbar$ 
(Ehrenfest) time scale.\par
We consider Wigner functions\cite{wigner,berry2} of a pure state $\ket{\psi}$
from the Hilbert space corresponding to a classical system with a compact phase space $\chi$. For explicit calculation we choose 2D toroidal phase space $\chi = \mathbb{T}_2$ that we quantize following Ref.~\citen{agam}. Let us write canonical Hilbert space basis of position states as $\ket{n}$, 
$n=0,\ldots,N-1$. Due to technical reasons $N$ is {\em odd}. 
WF is defined on a regular grid of $N\times N$ phase-space points
$$
  W_\psi(n,m)= \frac{N}{\sqrt{N-1}} \sum_{n',l} 
  e^{-2{\rm i}\pi n' m /N} \tilde \delta_ {2l-2n+n'} 
  \scalar{l+n'}{\psi}\scalar{\psi}{l},\quad 
  \tilde \delta_k= \frac{1}{N} \sum_{m'} e^{{\rm i} \pi m' k /N},
$$
where all indices run from  $-(N-1)/2$ to $(N-1)/2$, as will be assumed 
whenever we refer to the quantized torus. We study WF value distribution 
$P(w)$ defined as
\begin{equation} \label{eq:distrib}
 P(w) = \frac{1}{V} \int_{\chi} \delta(w-W(x)) dx,
\end{equation}
where $V$ is the phase space volume (area). 
In our case: $(1/V)\int_\chi = (1/N^2)\sum_{m,n}$. 
Let $\mean{(\ldots)}$ denote the average with respect to 
$P(w)$. The first two moments of $P(w)$, the mean value $\mean{W}$ and the standard deviation $\sigma$, 
are determined by the normalization of probability and purity $\rho^2=\rho$ 
of the state $\rho = \ket{\psi}\bra{\psi}$. 
We shall `normalize' WF by unitarizing the standard deviation
\begin{equation}
\sigma^2 = 1  \quad {\rm and} \quad \mean{W} = (N-1)^{-1/2}.
\label{eq:mom1}
\end{equation}
Particularly interesting is the {\em relative standard deviation} defined as $\kappa = \sigma/\mean{W}$ 
that scales with dimension $N$ as $\kappa \sim \sqrt{N}$, i.e. it {\it diverges} in the semi-classical limit.\par
The states of developed quantum chaos are associated with ergodic or random wave functions 
\cite{berry1} that can be defined in Hilbert space basis $\ket{l}$, $l=0,\ldots,N-1$ as
\begin{equation}
  \ket{\psi} = \sum_{l=1}^N c_l \ket{l},\qquad  \sum_{l=1}^N |c_l|^2 =1,
\label{eq:randomstate}
\end{equation}
where $c_l\in\mathbb{C}$ are random coefficients, that 
may be considered for large $N$ as independent 
Gaussian variables\cite{haake} of variance
 $\ave{|c_l|^2} = 1/N$. $\ave{\ldots}$ 
denotes ensemble average over random states (\ref{eq:randomstate}).
WF value statistics of a random state at fixed point in phase space should be transitionally invariant
and thus equal to the expected statistics of a `chaotic Wigner function'.
Therefore we can compute $P(w)$ by analyzing the distribution of WF values at one phase space point, e.g
$W(0,0)$,
\begin{equation}\label{eq:Wpoint}
  W(0,0) = \frac{N}{\sqrt{N-1}}
    \left( 
      |c_0|^2 + \sum_{l\neq l'} \tilde\delta(l+l') c^*_l c_{l'}
    \right).
\end{equation}
The expression (\ref{eq:Wpoint}) is a sum of $N$ statistically independent 
terms, where the first $|c_0|^2$ is strictly non-negative and others have 
vanishing mean and finite fluctuation. The central limit theorem, 
for large $N\gg 1$, implies that the distribution of $W(0,0)$ is a Gaussian
\begin{equation}
P_{\rm random}(w) = \frac{1}{\sqrt{2\pi}}\exp\left(-\frac{1}{2}(w-\mean{W})^2\right),
\label{eq:gaussian}
\end{equation}
with the first and the second moment equal to those of WF value distribution over the whole phase space (\ref{eq:mom1}).\par
\begin{figure}[htb]
  \begin{center}
    \includegraphics[angle=-90, width=12cm]{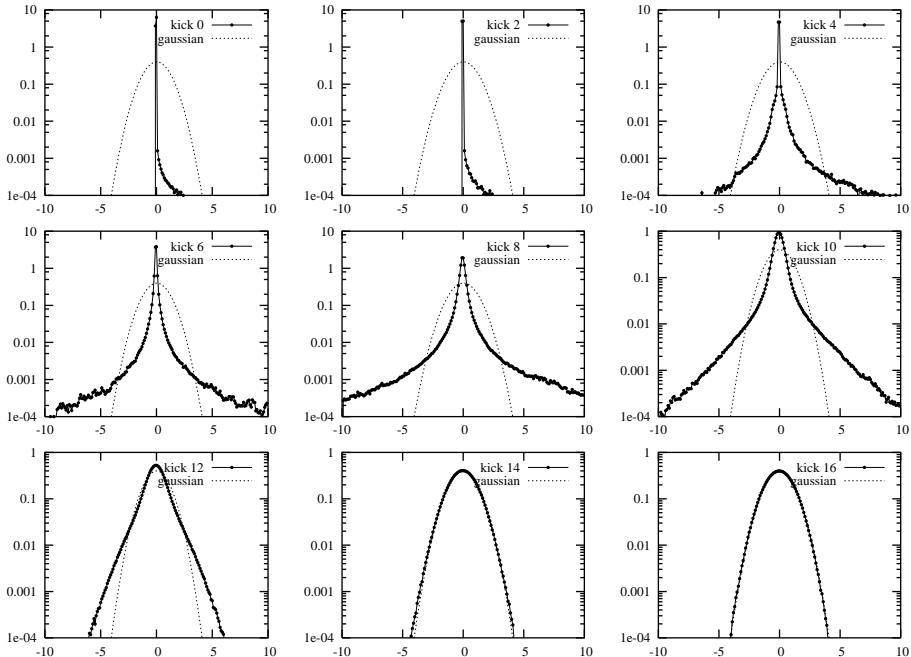}
  \end{center}
  \caption{Time evolution of the WF value distribution (\ref{eq:distrib}) of the quantized sawtooth map for $K=0.5$ and $N=2187$. Initial state is a coherent wave packet. Dashed curves give a theoretical Gaussian (\ref{eq:gaussian}).}
\label{pic:saw_stat}
\end{figure}

The dynamics is studied on the generic example of the quantized saw-tooth map on the toroidal phase space. The latter is with kicking strength $K$ defined by the following evolution operator\cite{casati}
$$
  U = \exp\left(-{\rm i}\frac{T}{2} {\hat p}^2\right)
  \exp\left({\rm i}\frac{K T}{2} {\hat n}^2\right),
$$
where $\hat n \ket{n}= n\ket{n}$, $\exp(2\pi{\rm i}\hat p/N)\ket{n}=\ket{n+1}$,
and $T= 2\pi/N$. Here we consider the regime $K>0$ for which the classical sawtooth map is ergodic and uniformly hyperbolic. In the fig.~\ref{pic:saw_stat} we show distribution $P(w)$ of time-evolving state in the sawtooth map for $K=0.5$ and $N=2187$, starting from a coherent state. We see that $P(w)$ quickly relaxes from a 
particular initial form to the Gaussian distribution 
(\ref{eq:gaussian}) predicted for random waves. 
The relaxation is the result of the wave self-interference caused by the 
corresponding classical stretching of the wave packet over the phase space.\par
\begin{figure}[htb]
\parbox{\halftext}{
  \begin{center}
    \includegraphics[angle=-90, width=6.5cm]{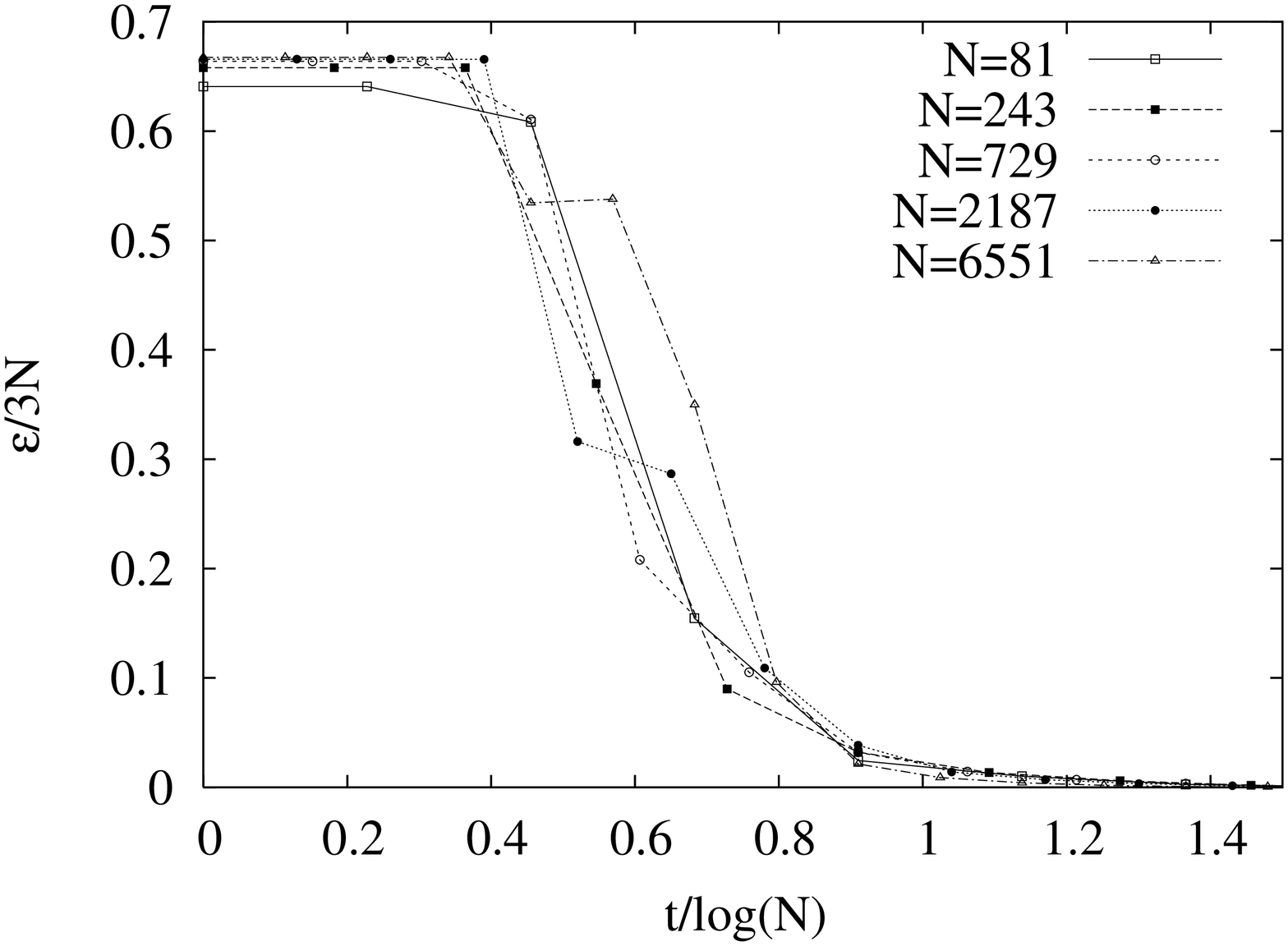}
  \end{center}
\caption{The excess $\epsilon$ of the WF value distribtion 
as a function of scaled time $t/\log N$ for several different values of $N$.
The parameter of the quantized sawtooth map is $K=0.5$, and the
initial state is a coherent wave packet.
}
\label{pic:saw_mom}}
\hfill
\parbox{\halftext}{
  \begin{center}
    \includegraphics[angle=-90, width=6.5cm]{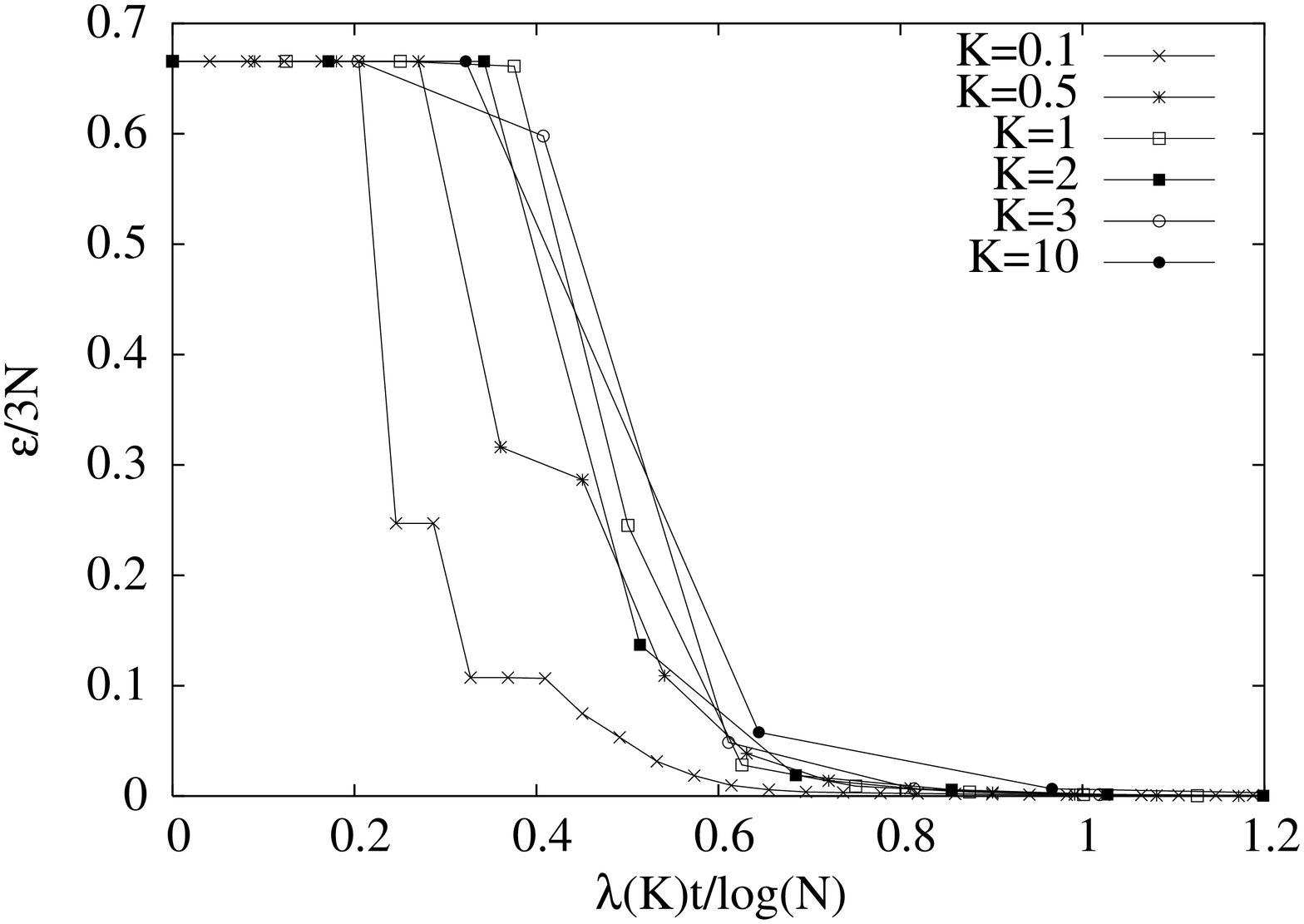}
  \end{center}
  \caption{Scaled time evolution of the excess $\epsilon$ for the 
quantized saw-tooth map at different kicking strengths $K$ 
and fixed dimension $N=2187$. The Lyapunov exponent is given by $\lambda(K) = \log((2+K + \sqrt{(2+K)^2-4})/2)$. \cite{casati}}
\label{pic:saw_cmp}}
\end{figure}

\begin{figure}\centering
    \includegraphics[angle=-90, width=6.5cm]{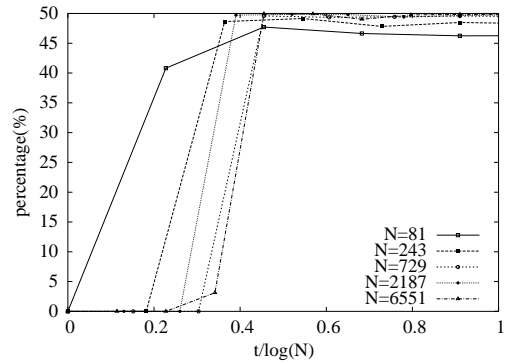}
 \caption{Percentage of phase space with negative WF as a function of scaled time for the quantized saw-tooth map, $K=0.5$, at several different dimensions $N$.
 }
\label{pic:saw_perc}
\end{figure}

To measure the deviation of distribution $P(w)$ from the Gaussian statistics we define the excess $\epsilon$ of the distribution by formula 
\begin{equation}
 \epsilon = \mean{(w-\mean{W})^4}/\sigma^4-3.
\label{eq:excess}
\end{equation}
Note, that $\epsilon = 0$ for a Gaussian (\ref{eq:gaussian}) and the size of $\epsilon$ measures a 
deviation from Gaussian. Using the excess $\epsilon$ we can quantitatively measure the transition of 
statistics to the asymptotic Gaussian form, that happens on the Ehrenfest time scale
\begin{equation}
  t_{\rm r} \sim \log(N)/\lambda,
\label{eq:trelax}
\end{equation}
where $\lambda$ is an effective Lyapunov exponent and $N$ is the dimension of Hilbert space. Numerical 
experiments for the family of sawtooth maps show that $t_{\rm r}$ (\ref{eq:trelax}) has the right 
scaling, both with $N$ (see fig.~\ref{pic:saw_mom}) and with $\lambda$ (see fig.~\ref{pic:saw_cmp}).
To check the rate of breakdown of direct classical-quantum correspondence we propose to study the 
percentage of phase space with negative valued WF 
\begin{equation}
  P_- =\int_{-\infty}^0 P(w) dw
\end{equation}
as a function of time. For random states $P_-$ goes to $1/2$ in the limit $N\to\infty$. 
In the fig.~\ref{pic:saw_perc} we show the evolution of $P_-$ in the quantized saw-tooth map, initially in 
a coherent state, for different dimensions $N$. We see that $P_-$ converges very fast to $1/2$ on a 
time scale $t_{\rm c}$ proportional to $\log(N)$ but significantly smaller than relaxation time 
$t_{\rm r}$. However, both time scales have the same scaling with $N$ and $\lambda$.\par
We have studied WF statistics of random states, and of time-dependent states of 
time-periodic, e.g. kicked quantum systems. 
The statistical approach to value statistics is meaningful since two-point 
phase space correlations of WF vanish (for distinct points) in the 
semi-classical limit as shown in a longer paper. \cite{HP01}
We have proven that random states corresponding to the regime of quantum chaos have Gaussian WF value 
statistics, with the relative standard deviation divided by the mean value diverging as 
$\sigma/ \mean{W} \sim \sqrt{N}$ in the semi-classical limit $N\to\infty$.
In addition we have shown that WF statistics in a classically fully chaotic system, initially in coherent state,
tends to a Gaussian statistics within the Ehrenfest time scale $t_{\rm r}$.
We believe that our results may shed some light onto the related studies of decoherence 
\cite{zurek,srednicki} and the fidelity of two initially equivalent states which undergo two 
slightly different time-evolutions. \cite{PZ02}\ 
The randomness of WFs in classically chaotic systems suggests a description 
by means random functions. 
The latter is the subject of the more detailed forthcoming publication \citen{HP01}.
	
\section*{Acknowledgments}
Useful discussions with A. B\" acker, G. Veble and M. \v Znidari\v c, as well as the financial support by the Ministry of Education, Science and Sport of Slovenia are gratefully acknowledged.

\end{document}